\documentclass[11pt]{article}
\usepackage[utf8x]{inputenc}
\usepackage[parfill]{parskip}
\usepackage{amsmath, amsfonts} 
\usepackage{graphicx}
\graphicspath{{images/}} 
\usepackage{subcaption}
\usepackage[a4paper, top=25mm, bottom=25mm, right=25mm, left=25mm]{geometry}
\usepackage{xcolor}
\usepackage{appendix}
\usepackage[noblocks]{authblk}
\usepackage[style=authoryear]{biblatex}
\bibliography{Bibiography}

\title{\bf Inference for Within- and Between-Partnership Transmission Rates for HIV Infection}
\author[1,*]{Irene García Muñoz}
\author[1]{Ian Hall}
\author[1]{Thomas House}
\affil[1]{Department of Mathematics, University of Manchester, Manchester, M13 9PL}
\affil[*]{Corresponding Author: irene.garciamunoz@manchester.ac.uk}
\begin{document}

\maketitle

\begin{abstract}
\noindent HIV transmission within serodiscordant couples remains a significant public health challenge, particularly in sub-Saharan Africa. Estimating the rate of such infection, alongside the rates of introduction of infection from outside the partnership, is a special case of the more general epidemiological challenge of inferring intensities of within- and between-group intensities of transmission.  This study presents a stochastic susceptible-infected (SI) pair model for estimating key epidemiological parameters governing HIV transmission within and between couples, which we further extend to account for gender-specific differences in infection dynamics. Using a likelihood-based inference approach, we estimate transmission parameters and associated uncertainty from observed data. These values can be used to inform infection prevention strategies for HIV, and the methodology proposed can be generalised to other epidemiological settings.

\end{abstract}

\section{Introduction}

In many infectious disease systems, transmission occurs through repeated contacts within closely connected sub-populations as well as through weaker contacts between these sub-populations, with examples being households \parencite{Lau:2012,Madewell:2020}, prisons \parencite{Beaudry:2020}, hospitals \parencite{LopezGarcia:2018}, and long-term care facilities \parencite{Nicolle:2000,Hall:2021}. For HIV, stable sexual partnerships are such a sub-population structure, and transmission is observed to occur via two routes -- within-partnership transmission and between-partnership transmission -- with the relative contribution of these routes to transmission having significant epidemiological and public health implications \parencite{Morris:1997,Eames:2004,Miller:2017}. Within-partnership spread reflects the risk of onward infection from an already infected partner, whereas between-partnership spread captures acquisition from the wider community. A detailed synthesis of information on HIV transmission rates and risk variables in this cohort can be found in the work by \textcite{eyawo2010hiv}. This meta-analysis has found important factors that influence the probability of transmission by combining data from several studies. Interventions such as antiretroviral therapy (ART) or pre-exposure prophylaxis (PrEP) often act more strongly on one route than the other, making it crucial to quantify each separately.

For HIV in sub-Saharan Africa, serodiscordant couples are a natural population in which to study these dynamics. They provide a direct setting where both within- and between-partnership events can be observed: internal transmission converts a discordant couple to a concordant positive, while external transmission converts a concordant negative couple to a discordant. Cohort studies tracking these transitions \parencite{hugonnet2002incidence, wall2019hiv, brahmbhatt2014longitudinal, shamu2021treatment} therefore offer an opportunity to estimate both forces of infection from empirical data.

In this study, we develop and apply a simplified pair-based modelling framework to a large retrospective cohort of stable heterosexual couples reported by \textcite{wall2019hiv}. Our goal is to infer the within-partnership transmission rate ($\tau$) and the between-partnership transmission rate ($\lambda$) directly from observed transitions between couple states. We do so in both a non-gendered model and an extended gender-specific model that distinguishes male-to-female and female-to-male internal transmission.

In terms of existing work, our model is related both to pair-formation models as reviewed by \textcite{Kretzschmar:2017} and differential-equation models for household epidemics as in Section 3 of \textcite{Ball:1999}, but simplified to allow closed-form solutions to be obtained. The likelihood takes the same approach as in \textcite{Kinyanjui:2018}, but using these exact solutions instead of numerical matrix exponentials. The identifiability of within- and between-group transmission rates was considered in \textcite{Ball:1997} using methods for the SIR case where recovery from infection leads to long-lasting immunity, and the data relate to the final outcome.

By combining exact analytical solutions of the model with maximum likelihood inference, we obtain precise estimates and uncertainty intervals for both rates $\tau$ and $\lambda$. This approach allows us to quantify the relative importance of internal and external transmission in sustaining HIV prevalence in stable couples, and to evaluate how these contributions vary when gender-specific effects are considered. The results have direct implications for intervention strategies, particularly for determining whether resources should prioritise blocking community acquisition or reducing within-couple spread.

\section{Methodology}

\subsection{Model description}

We developed a modelling framework to describe the transmission dynamics of a disease within a closed population in which every individual remains in a pair throughout the study. Since our focus is HIV, we use two epidemiological compartments: $S$ (susceptible) and $I$ (infected). Considering the state of a pair, we identify three configurations: $SS$ (both susceptible or seroconcordant  negative), $SI$ (one susceptible, one infected or serodiscordant), and $II$ (both infected  or seroconcordant  positive), with the model’s dynamical variables being the expected counts of each pair type. Transitions between these compartments are driven by two rates: the internal transmission rate $\tau$, accounting for infection within pairs, and the external transmission rate $\lambda$, representing infection from sources outside the pair.

In terms of the simplifications made to a typical sexually transmitted infection (STI) pair-formation model \parencite{Kretzschmar:2017}, the timescales over which we are modelling do not require explicit pairing dynamics such as partnership duration, monogamy, and intra-pair reinfection that are more epidemiologically important over longer periods. However, the core structure—tracking $SS$, $SI$, and $II$ pair states and using within-pair transmission ($\tau$) and external force of infection ($\lambda$)—remains faithful to their theoretical foundation.

The choice of pair-based modelling over classical models (which assume instantaneous partnerships and characterize transmission via partner-change rates) is supported by e.g.\ \textcite{ong2012comparability}, who compared both approaches across several STIs. They found that classical models can underestimate self-sustaining transmission in longer partnerships. This validates our use of the pair model, since it better captures HIV's extended infectious periods within partnerships while avoiding unnecessary model complexity.

A system of ordinary differential equations (ODEs) describes the model's dynamics and guarantees the conservation of total pairs throughout time.
\begin{align}
    \frac{\mathrm{d}P_{SS}}{\mathrm{d}t} & = -2\lambda P_{SS},\nonumber \\ \nonumber
    \frac{\mathrm{d}P_{SI}}{\mathrm{d}t} & = 2\lambda P_{SS} - (\lambda + \tau) P_{SI}, \\
   \frac{\mathrm{d}P_{II}}{\mathrm{d}t} & = (\lambda + \tau) P_{SI}. \label{eq: non-gender pairs}
\end{align}
These equations ensure the conservation of total pairs, that is, $P_{SS}+P_{SI}+P_{II} = N$, where $ N$ is the total number of pairs and involves 2 unknown parameters. 

Solving this system is possible, such that:
\begin{align*}
    P_{SS}(t) &= P_{SS}(0) {\rm e}^{-2\lambda t}; \\
    P_{SI}(t) &= \left(P_{SI}(0) {\rm e}^{-(\tau-\lambda) t} + P_{SS}(0)\frac{2\lambda}{(\tau -\lambda)} ( 1 - {\rm e}^{-(\tau-\lambda ) t}) \right){\rm e}^{-2\lambda t}.
\end{align*}
Full derivations of these and other mathematical results are given in the Supplementary Material. Note that we will use $P$ to refer to modelled expected numbers of a given pair state, which are continuous, and $N$ to refer to observed integer numbers of pairs, throughout.
Because the sum $P_{SS} + P_{SI} + P_{II} = N$ is conserved by Equations \eqref{eq: non-gender pairs}, we then have
\begin{equation*}
    P_{II}(t) =N - \left(P_{SS}(0)+P_{SI}(0) {\rm e}^{-(\tau-\lambda ) t} + P_{SS}(0)\frac{2\lambda}{\tau-\lambda} ( 1 - {\rm e}^{-(\tau-\lambda ) t}) \right){\rm e}^{-2\lambda t}.
\end{equation*}
It is worth noting here that $\tau$ appears in combination with $\lambda$ in the above system of equations and so it is likely worth defining $\tau=\lambda+\theta$ where $\theta$ is the additional transmission potential from internal contacts compared with external force of infection.

We can track male and female partners more explicitly as in \textcite{brahmbhatt2014longitudinal}. We make a distinction between those serodiscordant pairs where the male or female is infected. We divide our population into two classes, $m$ (male) and $f$ (female), such that
\begin{align*}
    \frac{{\rm d}P_{S_mS_f}}{{\rm d}t} &= - (\lambda_m + \lambda_f) P_{S_mS_f} , \\
    \frac{{\rm d}P_{I_mS_f}}{{\rm d}t} &= \lambda_m P_{S_mS_f} - (\tau_{m \rightarrow f} + \lambda_f) P_{I_mS_f}, \\
    \frac{{\rm d}P_{S_mI_f}}{{\rm d}t} &= \lambda_f P_{S_mS_f} - (\lambda_m + \tau_{f \rightarrow m}) P_{S_mI_f}, \\
    \frac{{\rm d}P_{I_mI_f}}{{\rm d}t} &= (\tau_{m \rightarrow f} + \lambda_f) P_{I_mS_f} + (\lambda_m + \tau_{f \rightarrow m}) P_{S_mI_f}.
\end{align*}
This model now has four unknown parameters. Again, we can solve this system exactly and obtain
\begin{align*}
    P_{S_mS_f}(t) &= P_{S_mS_f}(0) {\rm e}^{-(\lambda_m + \lambda_f) t} ,\\
    P_{I_mS_f}(t) &= \left(P_{I_mS_f}(0) {\rm e}^{-(\tau_{m \rightarrow f} - \lambda_m)t} + \frac{\lambda_m P_{S_mS_f}(0)}{\tau_{m \rightarrow f} - \lambda_m} (1 - {\rm e}^{-(\tau_{m \rightarrow f} - \lambda_m) t}) \right) {\rm e}^{-(\lambda_m + \lambda_f) t}, \\
    P_{S_mI_f}(t) &= \left(P_{S_mI_f}(0) {\rm e}^{-(\tau_{f \rightarrow m} - \lambda_f)t} +  \frac{\lambda_f P_{S_mS_f}(0)}{\tau_{f \rightarrow m} - \lambda_f} (1 - {\rm e}^{-(\tau_{f \rightarrow m} - \lambda_f) t}) \right) {\rm e}^{-(\lambda_m + \lambda_f) t},\\
    P_{I_mI_f}(t) &= N - {\rm e}^{-(\lambda_m + \lambda_f) t} \Bigg( P_{S_mS_f}(0) + P_{I_mS_f}(0) {\rm e}^{-(\tau_{m \rightarrow f} - \lambda_m)t} \\
    &\quad + P_{S_mI_f}(0) {\rm e}^{-(\tau_{f \rightarrow m} - \lambda_f)t} 
    + \frac{\lambda_m P_{S_mS_f}(0)}{\tau_{m \rightarrow f} - \lambda_m} \big(1 - {\rm e}^{-(\tau_{m \rightarrow f} - \lambda_m) t}\big) \\
    &\quad + \frac{\lambda_f P_{S_mS_f}(0)}{\tau_{f \rightarrow m} - \lambda_f} \big(1 - {\rm e}^{-(\tau_{f \rightarrow m} - \lambda_f) t}\big) \Bigg).
    \nonumber
\end{align*}

\subsection{Data}

We base our initial conditions and parameter estimation on data reported in a retrospective cohort study conducted in Mwanza, Tanzania, which followed $N=1,802$ stable heterosexual couples over a two-year period \parencite{hugonnet2002incidence}. The study provided detailed information about the HIV status of both partners at two time points, enabling us to track the transitions between different epidemiological states.

At the start of the study, $N_{SS}^0 = 1,742$ couples were identified as concordant negative, with both partners being HIV-negative. $N_{SI}^0 = 43$ couples were serodiscordant, $N_{I_m S_f}^0 = 22$ of these involved an HIV-positive male with a susceptible female partner, while $N_{S_mI_f}^0 = 21$ involved an HIV-positive female with a susceptible male. $N_{II}^0 = 17$ couples were concordant positive, where both partners were already infected. These values were used to set the initial conditions of the model, so $P_{SS}(0) = P_{S_mS_f}(0) = N_{SS}^0$ etc.

Two years later, at time $T$, the cohort was reassessed. The number of concordant-negative couples decreased to $N_{SS}^T = 1,721$. Meanwhile, the number of serodiscordant couples increased to $N_{SI}^T = 58$, composed of $N_{I_m S_f}^T = 33$ couples in which the male was infected and $N_{S_mI_f}^T = 25$ in which the female was infected. The number of concordant-positive couples rose to $N_{II}^T = 23$. These changes reflect a combination of internal transmission within couples and potential external acquisition of HIV, and they provide a basis for estimating both within-partnership (internal) and community-based (external) transmission parameters.

All model parameters and initial conditions are summarized in Table \ref{tab: parameters and data}. The total number of couples is fixed at $N=1,802$, and parameters such as $\lambda$ (external transmission rate) and $\tau$ (internal transmission rate) are initialized in numerical fitting based on plausible estimates from \textcite{wall2019hiv}, then evolved towards maximum-likelihood values using optimisation routines from \textit{scipy} \parencite{2020SciPy-NMeth}. Gender-specific versions of these parameters, $\lambda_m$, $\lambda_f$, $\tau_m$, and $\tau_f$, capture the directional aspects of HIV transmission observed in the cohort.

\begin{table}[ht!]
\centering
\begin{tabular}{|c|p{6cm}|p{6cm}|}
\hline
\textbf{Symbol} & \textbf{Description} & \textbf{Value / Notes} \\
\hline
$N_{SS}^t$ & Concordant negative (SS) pairs & 1742 at $t=0$, 1721 at $t=2$ \parencite{hugonnet2002incidence}\\
\hline
$N_{SI}^T$ & Serodiscordant (SI) pairs (1 susceptible, 1 infected) & 43 at $t=0$, 58 at $t=2$  \parencite{hugonnet2002incidence}\\
\hline
$N_{II^T}$ & Concordant positive (II) pairs & 17 at $t=0$, 23 at $t=2$  \parencite{hugonnet2002incidence}\\
\hline
$N_{I_mS_f}$ & Male infected, female susceptible pairs & 22 at $t=0$, 33 at $t=2$  \parencite{hugonnet2002incidence}\\
\hline
$N_{S_mI_f}$ & Female infected, male susceptible pairs & 21 at $t=0$, 25 at $t=2$ \parencite{hugonnet2002incidence}\\
\hline
$N$ & Total number of pairs (constant over time) & 1802 pairs  \parencite{hugonnet2002incidence}\\
\hline
$\lambda$ & External transmission rate &  Estimated \\
\hline
$\lambda_m$ & External transmission rate affecting male partners & Estimated \\
\hline
$\lambda_f$ & External transmission rate affecting female partners & Estimated \\
\hline
$\tau$ & Internal transmission rate (generic) & Estimated \\
\hline
$\tau_{m \rightarrow f}$ & Internal transmission rate from infected male to female & Estimated \\
\hline
$\tau_{f \rightarrow m}$ & Internal transmission rate from infected female to male & Estimated\\
\hline
\end{tabular}
\caption{Model Parameters and Data}
\label{tab: parameters and data}
\end{table}

\subsection{Parameter estimation}

The multinomial probability distribution is used to link the model outputs to the observed data in order to estimate the parameters $\lambda$ and $\tau$. Since we seek to provide methods that are generalisable, we will work with the generic case where data consists of arrays of observed counts $(N_{SS}^t, N_{SI}^t, N_{II}^t)$ for every observation time point $t \in \mathcal{T}$, where $\mathcal{T}$ is a general set of observation times and equal to $\{0, T\}$ for our specific dataset.
Then the likelihood function is provided by
\begin{equation*}
    L(\lambda, \tau) = \prod_{t\in \mathcal{T}} \binom{N^t}{N_{SS}^t, N_{SI}^t, N_{II}^t} P_{SS}(t)^{N_{SS}^t} P_{SI}(t)^{N_{SI}^t} P_{II}(t)^{N_{II}^t},
\end{equation*}
where \(P_{SS}(t)\), \(P_{SI}(t)\), and \(P_{II}(t)\) are the proportions predicted by the model given by Equations~\eqref{eq: non-gender pairs} at time \(t\), and \(N\) is the total number of pairs observed at that time point. Thus, the logarithm of this likelihood function simplifies to:

\begin{equation*}
    \mathcal{L}(\lambda, \tau) = \log(L(\lambda, \tau)) \propto \sum_{t\in \mathcal{T}} \big( N_{SS}^t \log(P_{SS}(t)) + N_{SI}^t \log(P_{SI}(t)) + N_{II}^t \log(P_{II}(t)) \big),
\end{equation*}
which forms the basis for parameter estimation, since the multinomial coefficients that form the constant of proportionality do not depend on the parameters to be estimated. The most likely values of $\lambda$ and $\tau$ given the observed data are obtained by maximising this log-likelihood function.

The curvature of the log-likelihood curve around its maximum is examined in order to measure the degree of uncertainty in the parameter estimates. This is accomplished by calculating the matrix of second partial derivatives of the negative log-likelihood, or the Hessian matrix:

\begin{equation*}
    \mathbf{H}(\lambda^*, \tau^*) = \left. \begin{bmatrix}
\frac{\partial^2 \mathcal{L}}{\partial \lambda^2} & \frac{\partial^2 \mathcal{L}}{\partial \lambda \partial \tau} \\
\frac{\partial^2 \mathcal{L}}{\partial \tau \partial \lambda} & \frac{\partial^2 \mathcal{L}}{\partial \tau^2}
\end{bmatrix}\right|_{(\lambda^*, \tau^*)}.
\end{equation*}

Standardly, if we have a maximum likelihood estimate for the parameters,
\begin{equation*}
\hat{\boldsymbol{\theta}}_{\mathrm{MLE}} = 
(\hat{\lambda}_{\mathrm{MLE}}, \hat{\tau}_{\mathrm{MLE}} )
= \mathrm{argmax} (\mathcal{L}(\lambda, \tau)) ,
\end{equation*}
then an asymptotic confidence region can be constructed from the contours of a multivariate normal probability density function with mean $\hat{\boldsymbol{\theta}}_{\mathrm{MLE}}$ and covariance matrix $-\mathbf{H}^{-1}(\hat{\lambda}_{\mathrm{MLE}}, \hat{\tau}_{\mathrm{MLE}} )$.

Finally, the identifiability and structure of the likelihood surface are explored by evaluating \(L(\lambda, \tau)\) over a grid of parameter values. This provides a visual representation of how well the parameters are resolved and reveals any correlations or ambiguities in their estimation.

\subsubsection{Likelihood Analysis: The non-gender model}

As mentioned above, we consider the special case $\mathcal{T} = \{0,T\}$. The likelihood involves six terms: three at time $0$ and three at time $T$, two years later. Only the three at time $T$ involve $\lambda$ or $\tau$ terms. At time $0$, we observe $N_{SS}^0$, $N_{SI}^0$, and $N_{II}^0$, and at time $T$ we observe $N_{SS}^T$, $N_{SI}^T$, and $N_{II}^T$. 
The log-likelihood involves constant terms that do not affect derivatives and is given by

\begin{equation*}
\begin{split}
    \mathcal{L} = N_{SS}^0 \log P_{SS}(T; \lambda) + N_{SI}^0 \log P_{SI}(T; \lambda, \tau) \\ + N_{II}^0 \log P_{II}(T; \lambda, \tau) +\mathrm{constant}.
\end{split}
\end{equation*}
Differentiating the log-likelihood with respect to a general parameter $\theta$ gives:
\begin{equation*}
    \frac{\partial L}{\partial \theta} =  \left(\frac{N_{SS}^0(N-P_{SI})-P_{SS}(N-N_{SI}^0)}{P_{SS}(T)(N-P_{SI}(T)-P_{SS}(T))}\right)\frac{\partial P_{SS}(T)}{\partial \theta} +
\left(\frac{(N-P_{SS}(T))N_{SI}^0-(N-N_{SS}^0)P_{SI}(T)}{P_{SI}(T)\left(N-P_{SI}(T)-P_{SS}(T)\right)}\right) \frac{\partial P_{SI}(T)}{\partial \theta}.
\end{equation*}
Differentiating the log-likelihood with respect to $\tau$ (remembering that $P_{SS}$ does not involve $\tau$) means
\begin{equation*}
    P_{SI}(T; \hat{\lambda},\hat{\tau})=N_{SI}^0\frac{(N  - P_{SS}(T;\hat{\lambda}))}{(N - N_{SS}^0)},
\end{equation*}
and substitution into the $\lambda$ partial derivative gives $N_{SS}^0/P_{SS}(T; \hat{\lambda})=N_{SI}^0/P_{SI}(T; \hat{ \lambda}, \hat{\tau})$, whence
\begin{equation}
    \hat{\lambda} = \frac{1}{2T} \log{\left(\frac{N_{SS}^0}{N_{SS}^T}\right)} .
    \label{Equ: Lambda hat}
\end{equation}
Now we can find $\hat{\tau}$ given $P_{SI}(T; \hat{\lambda},\hat{\tau}) = N_{SI}^0$ so:
\begin{equation*}
    \left(N_{SI}^T {\rm e}^{-(\hat{\tau} - \hat{\lambda}) T} +  \frac{N_{SS}^T\log \left(\frac{N_{SS}^T}{N_{SS}^0} \right)}{T } \frac{\left(1 - {\rm e}^{-(\tau - \lambda) T} \right)}{(\tau - \lambda)} \right) = \frac{N_{SI}^0 N_{SS}^T}{N_{SS}^0}.
\end{equation*}
Now, there is a choice of how to solve for $\tau$. Here we reparameterise so that $\tau = (2\phi + 1) \lambda$ and hence 
\begin{equation*}
    \left(\frac{N_{SI}^T}{N_{SS}^T}\phi-1\right) \left(\frac{N_{SS}^0}{N_{SS}^T}\right)^\phi   = \left(\frac{N_{SI}^0 \phi}{N_{SS}^0}-1\right).
\end{equation*}
 To obtain $\tau$ we can rewrite the term $\left(\frac{N_{SS}^0}{N_{SS}^0}\right)^{\phi}$ as:
\begin{equation*}
    \left(\frac{N_{SS}^0}{N_{SS}^0}\right)^{\phi} = \left(1 - \frac{N_{SS}^0 - N_{SS}^0}{N_{SS}^0}\right)^{\phi}\approx 1 - \phi \frac{N_{SS}^0 - N_{SS}^0}{N_{SS}^0} +\ldots
\end{equation*}
So for small $\frac{N_{SS}^T - N_{SS}^0}{N_{SS}^T}$, we can approximate via binomial expansion and the equation becomes:
\begin{equation*}
    \frac{\phi N_{SI}^0}{ N_{SS}^0} - 1 = \left(\frac{\phi N_{SI}^T}{ N_{SS}^T} - 1\right) \left(1 - \phi \frac{N_{SS}^T - N_{SS}^0}{N_{SS}^T}\right).
    \label{tau derivation equation}
\end{equation*}
Then
\begin{equation*}
\hat{\phi}   \approx \frac{N_{SS}^T}{(N_{SS}^T - N_{SS}^0)}\left(  \frac{N_{SI}^0N_{SS}^T}{ N_{SS}^0N_{SI}^T} - 1  - \frac{ (N_{SS}^T - N_{SS}^0)}{N_{SI}^T}  \right).
\end{equation*}
Finally, we can use the reparameterization $\tau = (2\phi + 1) \lambda$ to estimate $\hat{\tau}$. We note that in general, $\hat{\lambda} = \hat{\lambda}_{\mathrm{MLE}}$, but $\hat{\tau} = \hat{\tau}_{\mathrm{MLE}} + \mathcal{O}(\phi^2)$. We will call these results `Analytical' and those obtained from a full numerical optimisation `MLE (refined)'.

\subsubsection{Likelihood Analysis: The gender model}
Following the same notation as above, at time $0$, we observe $N_{SS}^0$, $N^0_{I_mS_f}$, $N_{S_mI_f}^0$, and $N_{II}^0$, and at time $T$ we observe $N_{SS}^T$, $N_{I_mS_f}^T$, $N_{S_mI_f}^T$, and $N_{II}^T$. 
The gender model can be re-parametrised in the following way:
\begin{align*}
\lambda_m &= 2\lambda q, \\
\lambda_f &= 2\lambda (1 - q), \\
\tau_{m \rightarrow f} &=  2\lambda (q + \theta_m), \\
\tau_{f \rightarrow m} &=  2\lambda (1 - q+\theta_f).
\end{align*}
Thus
\begin{align*}
    P_{S_mS_f}(t) &= N_{SS}^t {\rm e}^{-2\lambda t}\\
    P_{I_mS_f}(t) &= \left(N_{I_mS_f}^t {\rm e}^{-2\lambda \theta_m t} + \frac{q N_{SS}^t}{ \theta_m} (1 - {\rm e}^{-2\lambda \theta_m t}) \right) {\rm e}^{-2\lambda t}. \\
    P_{S_mI_f}(t) &= \left(N_{S_mI_f}^t {\rm e}^{-2  \lambda\theta_ft} +  \frac{(1-q)) N_{SS}^t}{\theta_f} \left(1 - {\rm e}^{-2\lambda\theta_f t}\right) \right) {\rm e}^{-2\lambda t}.
\end{align*}
The log-likelihood without constant terms that do not affect derivatives therefore reduces to:
\begin{equation*}
    \mathcal{L} = N_{SS}^T \log P_{SS}(T) + N_{I_mS_f}^T \log P_{IS}(T) + N_{S_mI_f}^T \log P_{SI}(T) + N_{II}^T \log P_{II}(T).
\end{equation*}
By an analogous argument to the non-gendered version we expect the MLE to occur when $P_{SS}(T;\hat{\lambda})=N_{SS}^0$
hence \[    e^{-2\hat{\lambda}T} = \left(\frac{N_{SS}^0}{N_{SS}^T} \right).\]
We can then derive an explicit form for $\theta_m$ and $\theta_f$ conditional on $\hat{q}$ and $\hat{\lambda}$ by same approach as above, giving
\begin{align*}
    \frac{\theta_m N_{I_mS_f}^0}{N_{SS}^0} -q&= \left(\frac{\theta_m N_{I_mS_f}^T}{N_{SS}^T}  -q  \right)\left(\frac{N_{SS}^0}{N_{SS}^T} \right)^{\theta_m}  \\
    \frac{\theta_f N_{S_mI_f}^0}{N_{SS}^0} -1+q&= \left(\frac{\theta_f N_{S_mI_f}^T}{N_{SS}^T}  -  (1-q)  \right) \left(\frac{N_{SS}^0}{N_{SS}^T} \right)^{\theta_f}
\end{align*}
and hence
\begin{align}
   \theta_m  &\approx \left(q+\left(  \frac{N_{SS}^T }{N_{SS}^T - N_{SS}^0} \right)\left(\frac{N_{I_mS_f}^T}{N_{SS}^T}   -\frac{ N_{I_mS_f}^0}{N_{SS}^0}  \right) \right)\frac{ N_{SS}^T}{N_{I_mS_f}^T}\\
   \theta_f  &\approx \left(1-q+\left(  \frac{N_{SS}^T }{N_{SS}^T- N_{SS}^0} \right)\left(\frac{N_{S_mI_f}^T}{N_{SS}^T} -\frac{ N_{S_mI_f}^0}{N_{SS}^0}  \right) \right)\frac{ N_{SS}^T}{N_{S_mI_f}^T}. \label{eq:gendertheta}
\end{align}
By deriving the `Analytical' results above, we can avoid numerical optimisation required to obtain `MLE (refined)' results in some cases, and for all models provide insight to the way in which different observations influence estimates and identifiability.

\section{Results}
\label{Results}

\subsection{Analytical Results}

For the non-gender version, the analytical results are straightforward: substituting values, for the external force of infection rate, we get:
\begin{equation*}
    \lambda = \frac{1}{4} \log \left(\frac{1742}{1721} \right) \approx 0.003.
\end{equation*}
And for the internal force of infection, we can use the reparameterization $\tau = (\phi + 1) \lambda$ to estimate $\hat{\tau}$:
\begin{equation*}
    \hat{\tau} = (16.95 + 1) \hat{\lambda} = 17.95 \hat{\lambda}  = 0.054
\end{equation*}
For the gender model, we get the same value for $\lambda$ as we did in the non-gender model. However, when solving \eqref{eq:gendertheta} for the gender model, we find that higher terms in the Taylor series are needed to obtain unique maxima, suggesting use of numerical optimisation becomes preferable to analytical results for this more complicated model.

\subsection{Numerical optimisation results}

For the pair model, the external transmission rate ($\lambda$) and internal transmission rate ($\tau$) estimations were calculated using the parameter inference methods mentioned. We start with an initial parameter estimate based on the seroincidence rates calculated as in \textcite{wall2019hiv}:
\begin{align*}
    \lambda & \approx (N_{SI}^T - N_{SI}^0/(2T N_{SS}^0)); \\
    \tau & \approx (N_{II}^T - N_{II}^0/(2T N_{SI}^0)).
\end{align*}
We call this pair of parameter values the `Closed Form Approximation' (CFA).

The maximum likelihood estimates are $\hat{\lambda}= 0.003$, and $\hat{\tau}= 0.056$. At the maximum likelihood point, the Hessian matrix of the negative log-likelihood function was calculated in order to evaluate the uncertainty of these estimates. The covariance matrix that was produced yielded standard errors of $\sigma_{\lambda}= 0.001$ and $\sigma_{\tau}= 0.046$.

\begin{figure}[h!]
    \centering
    \includegraphics[width=\textwidth]{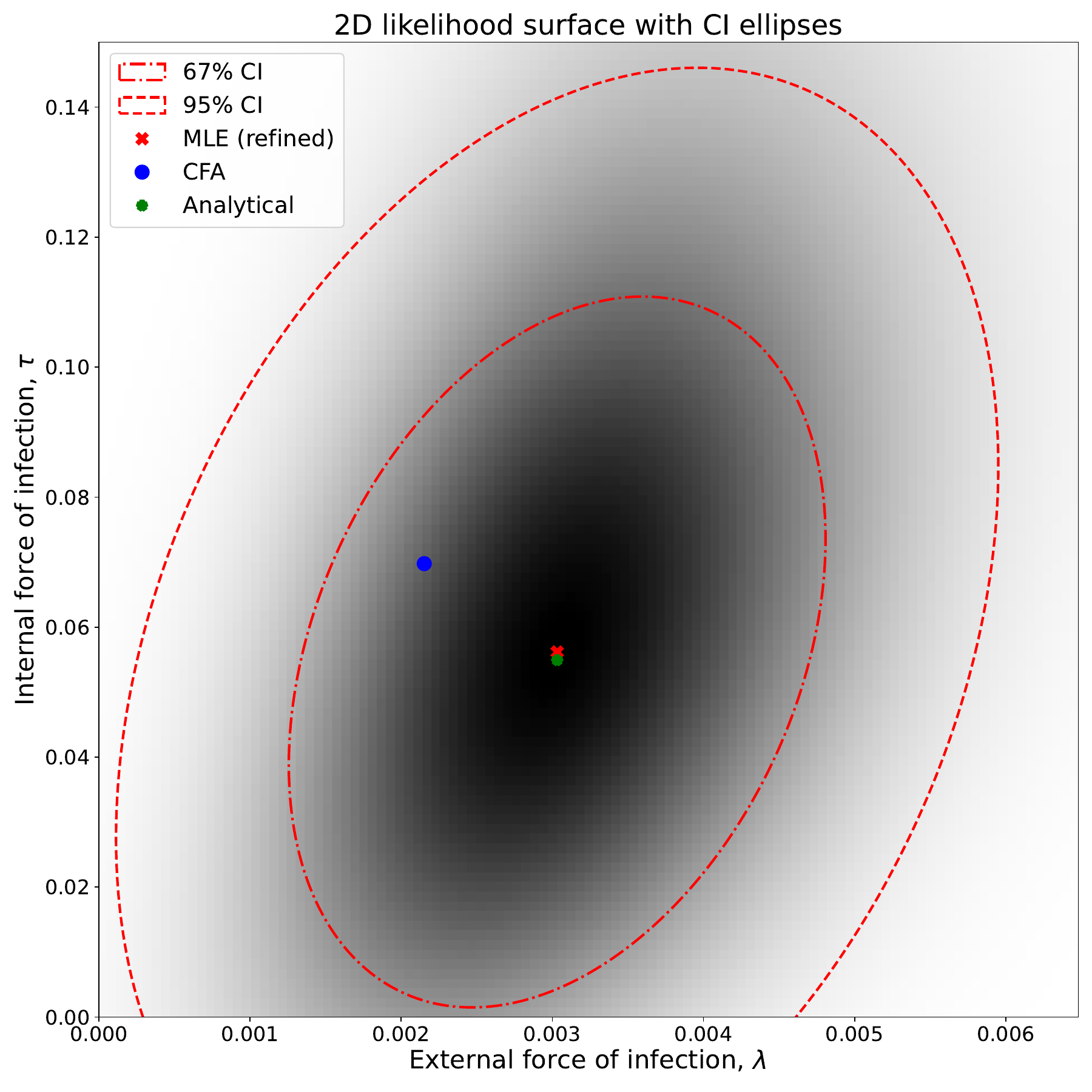}
    \caption{2D likelihood surface of the external ($\lambda$) and internal ($\tau$) forces of infection. The plot displays the maximum likelihood estimate (MLE), the closed-form approximation (CFA), and analytical estimates, along with 67\% and 95\% confidence interval (CI) ellipses. The CFA provides a quick estimate derived from a deterministic simplification of the underlying stochastic model.}
    \label{fig:SI pair heatmap}
\end{figure}

Using the standard errors, the $95\%$ and $67\%$ bivariate confidence regions for the parameters were created, shown in Figure \ref{fig:SI pair heatmap}. These intervals show that the observed data has a good resolution of the model parameters and comparison with the explicit likelihood surface shows that the standard approximation to the Hessian is justified provided the associated Gaussian is truncated at $\lambda$ and $\tau$ equal to $0$.

We now consider how parameters can be interpreted as infections per year in the population analysed.
Based on the initial number of couples in the model, $1,742$ susceptible–susceptible pairs ($N_{SS}^0$) and $43$ discordant pairs ($N_{SI}^0$), we can estimate the number of new infections per year due to both internal and external forces of infection. Using an external infection rate $\lambda = 0.003$, and considering that there are 3,527 susceptible individuals (two per SS pair and one per SI pair), we estimate approximately 11 new infections per year from external sources.

For internal transmission, where infection occurs within discordant couples, the number of new infections per year is given by $\tau \times 43$, where $\tau = 0.056$ is the internal force of infection, leading to around 2.4 new infections per year from within-partnership transmission. 

\begin{table}[h!]
\centering
\begin{tabular}{|l|c|c|}
\hline
\textbf{Transmission Type} & \textbf{Estimated Infections/Year} & \textbf{Per 1,000/year}\\
\hline
External ($\lambda = 0.00303$) & 10.6 & 3.03\\
\hline
Internal ($\tau = 0.0561$) & 2.48 & 56.1\\
\hline
\textbf{Total} & \textbf{13.1} & \textbf{59.1}\\
\hline
\end{tabular}
\label{tab:non_gender_infections}
\caption{Estimated annual number of HIV infections calculated using non-gendered transmission rates. Infections are categorised as arising from external sources (community transmission) or internal sources (within-couple transmission).}
\end{table}

Similarly, for the male-female pair model, the external transmission rate ($\lambda$) and internal transmission rate ($\tau$) estimations were calculated using the parameter inference methods mentioned. Using the maximum likelihood estimates, $\hat{\lambda}_m = 0.004$, $\hat{\lambda}_f = 0.002$, $\hat{\tau}_{m \rightarrow f}= 0.047$ and $\hat{\tau}_{f \rightarrow m}= 0.068$. At the maximum likelihood point, the Hessian matrix of the negative log-likelihood function was calculated in order to evaluate the uncertainty of these estimates.

The $95\%$ confidence intervals for the parameters were generated using the standard errors, but with truncation at $0$, generating $(0.0006,0.0073)$ for $\lambda_m$, $(0.00,0.0051)$ for $\lambda_f$, $(0.00,0.1819)$ for $\tau_{m \rightarrow f}$ and $(0.00, 0.2271)$ for $\tau_{f \rightarrow m}$. These intervals demonstrate that the model parameters are reasonably informed by the observed data.

\begin{table}[h!]
\centering
\begin{tabular}{|l|c|c|}
\hline
\textbf{Transmission Type} & \textbf{Estimated Infections/Year} & \textbf{Per 1,000/year}\\
\hline
External ($\lambda_m = 0.004$) & 7.05 & 3.95\\
\hline
External ($\lambda_f = 0.002)$ & 3.52 & 2.11\\
\hline
Internal ($\tau_{m \rightarrow f} = 0.047$) & 1.03 & 46.5 \\
\hline
Internal ($\tau_{f \rightarrow m} = 0.068$) & 1.45 & 67.9 \\
\hline
\textbf{Total} & \textbf{13} & \textbf{120}\\
\hline
\end{tabular}
\label{tab:gender_infections}
\caption{Estimated annual number of HIV infections using gender-specific transmission rates. External infections affect susceptible individuals through community transmission, while internal infections occur within serodiscordant couples, from male to female and vice versa.}
\end{table}

Furthermore, the likelihood surface was assessed across a grid of $\lambda$ and $\tau$ values, resulting in a heat map that depicts the parameter space's structure. The visualisation verified that the parameters are distinguishable and the likelihood surface is unimodal, with the highest likelihood region's centre housing the maximum likelihood estimates.

\begin{figure}[ht!]
    \centering
    \includegraphics[width=\textwidth]{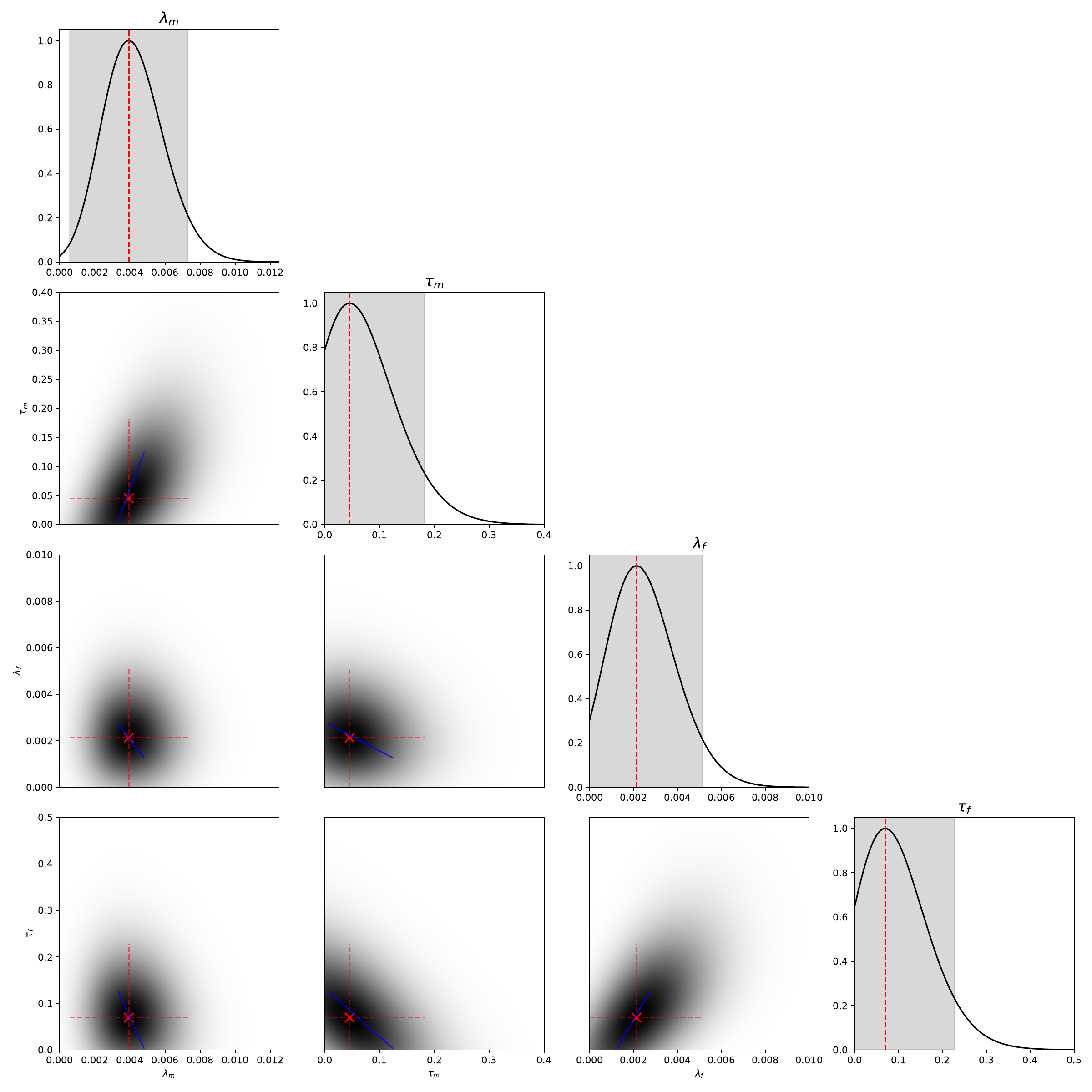}
    \caption{Each diagonal panel shows the profile likelihood curve for one parameter, computed by varying that parameter while fixing the others at their maximum likelihood estimates (MLE). The red dashed line indicates the MLE, and the grey shaded region corresponds to the $95\%$ confidence interval. The lower triangle panels display heatmaps of the joint likelihood (on the original likelihood scale) for each pair of parameters, again holding the remaining parameters at their MLEs. These last also show a blue line for the asymptotic solution.}
    \label{fig:SI mf pair heatmap}
\end{figure}

\subsection{Parameter Estimation Performance}

The estimation of $\lambda$ (the external transmission rate) and $\tau$ (the internal transmission rate) was evaluated over varying true parameter values to assess the robustness and accuracy of the inference framework. The performance was analysed using simulated datasets that reflect the realistic dynamics of serodiscordant couples.

\begin{figure}[htbp]
    \centering
    \includegraphics[width=\textwidth]{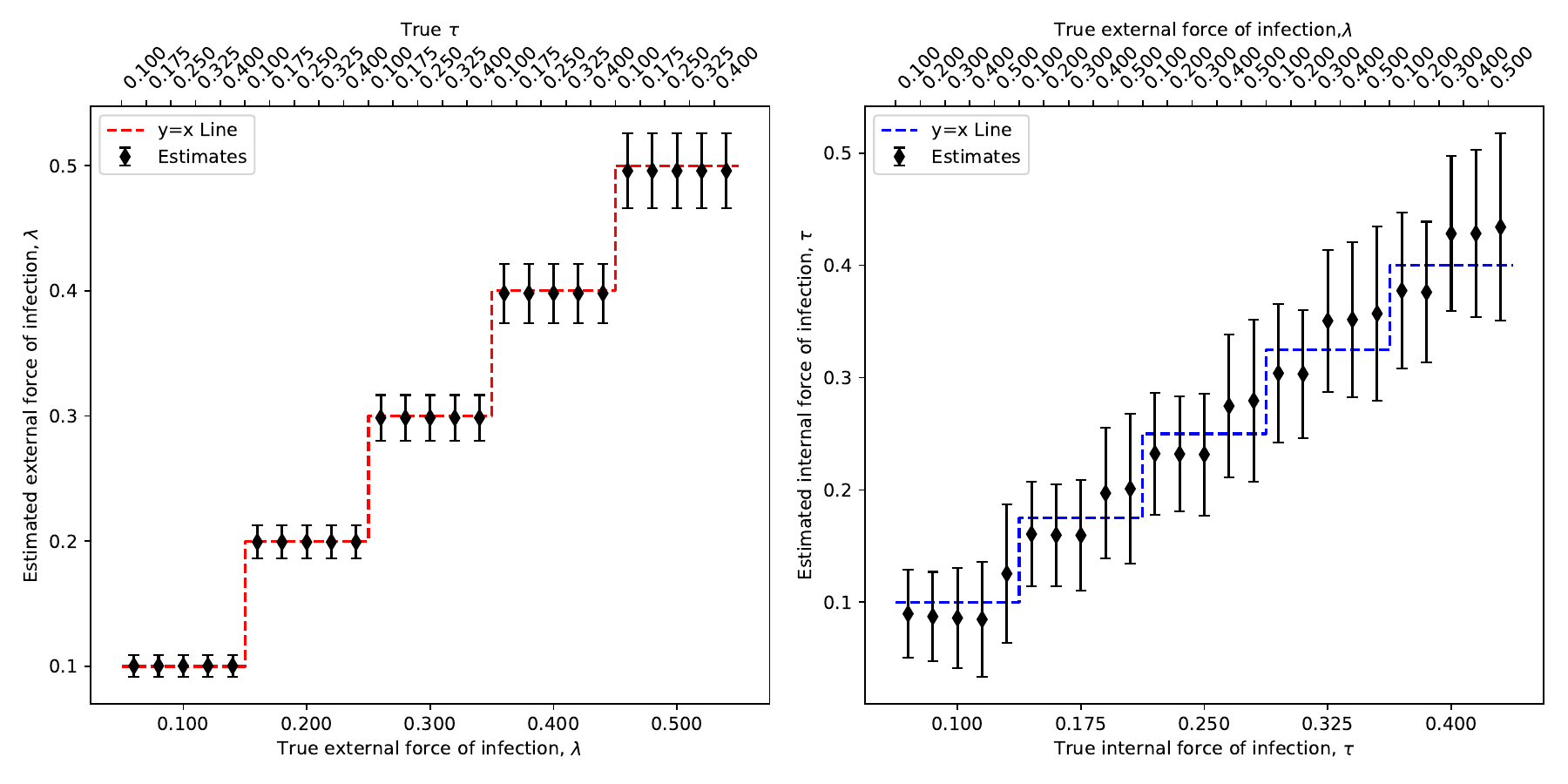}
    \caption{Validation of parameter estimation performance. The estimated external transmission rate ($\lambda$) and internal transmission rate ($\tau$) are plotted against their true values. Each point represents a simulated dataset, and the solid diagonal line corresponds to the identity line ($y = x$), indicating perfect estimation. The close alignment of points to the diagonal demonstrates the accuracy and reliability of the inference framework across a range of parameter values.}
    \label{fig:parameter range validation}
\end{figure}

Figure \ref{fig:parameter range validation} presents the estimated parameters plotted against their respective true values. The robust performance of the parameter inference framework highlights its applicability to real-world datasets. By accurately resolving key epidemiological parameters, the method offers valuable insights into transmission dynamics.

\section{Discussion}

The primary contribution of this paper is to formalise a within–between transmission decomposition for HIV in serodiscordant couples through a tractable stochastic pair model with precise solutions. Addressing a problem in HIV epidemiology, separating acquisition from the community from onward transmission within stable partnerships using cohort data that only observes pair states at discrete times. Our framework allows for both analytical and numerical approaches, enabling a transparent likelihood construction and closed-form approximate MLEs, which can generate insights both for the HIV data we consider here as well as for analogues in other settings like households, care homes, prisons and other sexual networks.

The structure of the non-gendered model is simple, yet it captures multiple susceptibles per pair (factor of 2 in $SS$), ongoing exposure of $SI$ pairs through both routes and the conservation of total pairs. The model does assume no pair dissolution, but it is a defensible position given the two year window of observation in the study, and it does mean the model estimates effective rates rather than biological per-act transmission probabilities. Another model assumption is the homogeneity of the external transmission, i.e. all susceptible individuals experience the same external force of infection ($\lambda$). This implies no behavioural heterogeneity and no dependence on the partner’s status. As a result, $\lambda$ should be interpreted as a background hazard dependent on the cohort relationship, not as a population-level incidence rate.

Using a multinomial likelihood conditional on total pairs is appropriate given that State counts are exhaustive, i.e. every pair in the study population is observed to be in exactly one and only one of the model’s states at each observation time; and that the sampling variability dominates over demographic stochasticity. From looking at the non-gender ODEs, they indicate that the likelihood is driven by $P_{SS}$, which identifies $\lambda$ and $P_{SI}$ and $P_{II}$, which identify $\tau$ conditional on $\lambda$. The result in Equation \ref{Equ: Lambda hat} shows that $\lambda$ is solely identifiable from the depletion of susceptibles, making it independent of $\tau$ and the SI pairs and II pairs counts. By contrast, $\tau$ is inferred from differences of small numbers, leading to wider confidence intervals and a strong curvature asymmetry in the likelihood. The estimation by reparameterization and binomial expansion is mathematically correct; yet, it exposes a more underlying concern: $\tau$ is only weakly identifiable from two time points unless the $SI$ counts exhibit significant variation.

It is observed that the internal force of infection $\tau$ appears only through combinations like $(\tau -\lambda)$, motivating the reparametrisation, $\tau =  \lambda + \theta$. It allows us to interpret $\lambda$, the external force of infection, as a baseline acquisition risk and $\theta$ as the excess risk attributable to having an infected partner. In the gendered model, the analogous decomposition $\tau_{m \rightarrow f} = \lambda_f + \theta_m$ and $\tau_{f \rightarrow m} = \lambda_m + \theta_f$ would further sharpen interpretation and may reduce posterior correlation.

The gender model introduces more structure, while remaining closely constrained by data availability. There are more identifiability constraints: two time points, four transition pathways and small $SI$ subgroups explain the wider confidence intervals, a dependence on the choice of parameter $q$ and the near ridges in the likelihood surface. This $q$ parameter is the proportion of the total external force of infection acting on men, which, epidemiologically speaking, we say $q$ captures gender asymmetry in community acquisition risk, not transmission efficiency. It reflects differences in background of prevalence in sexual networks, partner concurrency, age-mixing, and behavioural exposure but not biological per-act transmissibility (that is included in $\tau_{m \rightarrow f}$ and $\tau_{f \rightarrow m}$. The parameter $q$ is also subject to an identifiability issue because the $SS$ dynamics contain no information about $q$. All information about $q$, therefore, comes from the relative growth of $P_{I_mS_f}$ and $P_{S_mI_m}$. Nonetheless, these compartments start small and change moderately, while simultaneously influenced by $\lambda$, $\theta_m$ and $\theta_f$. Hence $q$ competes with $\theta_m$ and $\theta_f$ in explaining the same data. There is no unique value for $\hat{q}$ supported by the likelihood unless you fix $\theta_m$ and $\theta_f$, or you introduce strong priors, or you add more observation times. It is important to highlight that this is not a failure of inference; but a correct reflection of data limitations. Despite this, the point estimates are plausible. A higher female to male internal transmission is consistent with empirical literature, and the external risk being higher for men aligns with known prevalence patterns.

There are of course key limitations to our approach. Real-world dynamics may contain other complexities that are not specifically covered by this framework, such as changes in partnerships or outside interventions or coexisting conditions as in \textcite{perez1998herpes} and \textcite{stein2019long}. We did not take into account partnership dissolution or ART uptake, and we assumed a homogeneous external force of infection. However, this research makes a substantive methodological contribution by showing that a pair based model can be analytically tractable, and we believe that our insight that
within– and between-group transmission can be inferred directly from cohort data is robust to such complexities if they can be appropriately captured in future work.

\section*{Code availability}

The code for this paper can be found at:
\noindent https://github.com/igmsb18/Inference-for-Within--and-Between-Partnership-Transmission-Rates-for-HIV-Infection \\

\section*{Acknowledgements}

IGM would like to acknowledge PhD funding from the University of Manchester and UK Health Security Agency. IH and TH are supported by the Wellcome Trust (Ref: 227438/Z/23/Z) and Medical Research Council (Ref: UKRI483). Authors would like to thank Maria Zambon and Joe Hilton for helpful comments on this work.

\printbibliography

\end{document}